\documentclass[%
 reprint,
nofootinbib,
 amsmath,amssymb,
 aps,
pre,
]{revtex4-1}

\usepackage{graphicx}
\usepackage{dcolumn}
\usepackage{bm}

\usepackage{color} 
\usepackage{tabularx}

\begin{document}

\preprint{APS/123-QED}

\title{Single-mode Nonlinear Langevin Emulation of Magnetohydrodynamic Turbulence}

\author{Riddhi Bandyopadhyay}
 \email{riddhib@udel.edu}
\author{William H. Matthaeus} %
\email{whm@udel.edu}
\author{Tulasi N. Parashar}%
\affiliation{%
Department of Physics and Astronomy and Bartol Research Institute, University of Delaware, Newark, Delaware, U.S.A.
}%


\date{\today}

\begin{abstract}
Based on the Langevin equation of Brownian motion, we present a simple 
model that emulates a typical mode in 
incompressible magnetohydrodynamic turbulence, providing a demonstration
of several key properties.
The 
model equation is consistent with von K\'arm\'an decay law and Kolmogorov's symmetries. We primarily focus on the behavior of inertial range modes although we also attempt to include some properties of the large scale modes. Dissipation scales are not considered. 
Results from the model are compared with results from published Direct Numerical Simulations.
\begin{description}
\item[PACS numbers]
May be entered using the \verb+\pacs{#1}+ command.
\end{description}
\end{abstract}

\pacs{Valid PACS appear here}
\maketitle

 
\section{Introduction}
Turbulence, whether in a hydrodynamic, magnetohydrodynamic (MHD) 
or weakly collisional plasma, is characterized by  
nonlinear effects that produce complex observable 
signatures. 
In strong turbulence 
the dynamics 
recorded in individual measurements 
become sufficiently complex that the 
underlying degrees of freedom are often 
treated as random variables.
This type of behavior would be 
expected, for example, 
in pointwise measurements of velocity,
in strong turbulence, and 
increasingly so 
at large mechanical or magnetic 
Reynolds numbers.
Much of turbulence theory proceeds
through a {\it statistical} 
characterization of this behavior.
Dissipation and linear effects such as waves 
may act to attenuate, organize, and modify the 
intensity of the nonlinear effects, thus 
modifying the statistical characterization.
Here we are concerned with the particular problem
of modeling statistics of 
magnetoydrodynamic turbulence in the presence of an 
externally supported mean magnetic field.  
We employ a simple model to demonstrate
stochastic behavior of an inertial range 
Fourier mode, while it also 
engages 
in wave like couplings associated with the mean magnetic field. 
Rather than attempting a formal analysis, 
we develop  a model starting with 
a simple stochastic differential equation, 
the Langevin equation. After
suitable 
modifications, we 
arrive at a model with two real degrees of freedom and a nonlinear 
damping force. The results validate the model design and  reproduces several basic features of the MHD turbulence cascade. We view the purpose of a simple model such as the present one as demonstrating salient statistical properties of MHD turbulence rather than either predicting or explaining physical effects. In this way the purpose parallels that of the Langevin formalism, which is well known to demonstrate properties of Brownian motion. 

\section{Background}

The Langevin equation provides a reasonable starting 
point to represent certain  
stochastic properties of turbulence 
and turbulent transport.  
A classical result \cite{Uhlenbeck:PR1930} 
is that 
the Langevin equation demonstrates
the relaxation of a particle with an initial peculiar velocity
to a state consistent with thermal 
equilibrium, and a Maxwellian (Gaussian) distribution.
It follows that the displacements represent an instance of 
Brownian motion. The statistics of solutions
to the Langevin equation
formally obey a Fokker Planck equation~\cite{Uhlenbeck:PR1930}, 
which links this 
approach to many possible applications in diffusion 
and turbulence theory.
It has been shown that 
even the 
linear damping Langevin formulation 
can form the basis for 
representation 
of the more complex physics
occurring in turbulence. 
For example, Beck~\cite{Beck:PRE1994}
showed that a linear Langevin formulation 
with chaotic forcing can describe the turbulence 
statistics of inertial range velocity increments.
Analysis of a linear Langevin model 
incorporating oscillations 
associated with an  
applied magnetic field
was carried out by Balescu \textit{et al.}~\cite{Balescu1994PoP}
for application to charged particle diffusion.
TenBarge \textit{et al.}~\cite{TenBarge:CPC2014} employed such a damped,
driven linear oscillator as a method for driving plasma simulation. 
These studies
represent antecedents to the approach and goals of the present work.

The random behavior of turbulent fields in real space is mirrored by 
random behavior of its Fourier modes. This relationship becomes direct 
in a simple spatially periodic model representation of 
homogeneous turbulence. 
Relying in each case on 
an assumption of ergodicity, Gibbsian statistical 
mechanics~\cite{Lee1952} 
has been fruitfully applied to the ideal, finite dimensional (truncated) Galerkin model
of both hydrodynamic 
and MHD turbulence~\cite{Kraichnan1973JFM, Frisch:JFM1975,Kraichnan:RPP1980}. 
The impressive accuracy of the Gibbsian predictions 
for various wavenumber spectra in these models provides
affirmative support to the assumption of ergodicity of the real and 
imaginary parts of the Fourier amplitudes of velocity 
and, where applicable, magnetic fields.\footnote{
Long wavelength modes sometimes exhibit {\it broken ergodicity}
~\cite{Shebalin:PD1989, Shebalin:GAFD2013} associated with degeneracy 
of very large scale degrees of freedom. Analysis of longer simulations
has led to the suggestion that over very long time scales, 
ergodicity is restored~\cite{Servidio:PRE2008a}.}

\begin{figure}
\begin{center}
\includegraphics[scale=1.0]{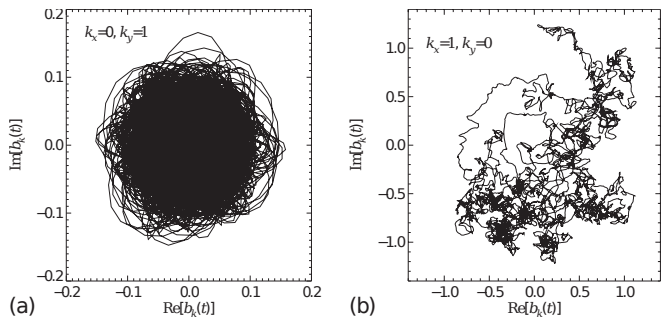}
\caption{Phase plane
trajectories over time of two Fourier modes. 
(left) Mode with a high degree of wave-like behavior; (right)
mode exhibiting very stochastic behavior and no indication of 
wave-like behavior. (Adapted from Dmitruk \textit{et al.}, 2009~\cite{Dmitruk:PoP2009})}
\label{fig:phaseplane}
\end{center}
\end{figure}

The stochastic behavior of individual Fourier degrees of freedom 
survives into the regime of driven, dissipative
high Reynolds number MHD turbulence e.g.,~\cite{Dmitruk:PoP2009}. 
However, 
when MHD turbulence evolves in the presence of an externally supported DC 
magnetic field, wave motions are induced, including the 
Alfv\'en wave, which survives to the incompressible limit.
Wave propagation itself does not 
cause stochastic behavior, 
but rather induces an orderly rotation
in the complex $(Re,Im)$ phase plane.
This contrast in behavior is illustrated in 
Fig. (\ref{fig:phaseplane}), which shows the time history, 
of the real and imaginary parts 
of the Fourier amplitude of a single cartesian magnetic field 
component, in a driven, 
dissipative MHD simulation~\cite{Dmitruk:PoP2009}.
It is apparent that one of these trajectories 
acts more like a constrained random walk, while the 
other admits a stronger sense of rotation as 
it is more dominantly  
influenced by the DC magnetic field. 
Demonstrating the balance and 
transition between these two types 
of behavior by varying control parameters for inertial range modes
is a major goal in the present paper. 

Stochastic behavior of a dynamical 
system motivates development of 
statistical theories that 
invoke ergodicity, e.g., the property that
permits replacement of ensemble averages by time averages.
The stochastic aspects of MHD or plasma 
dynamics might become less prominent
for certain modes
when an applied magnetic field of sufficient strength is present.
In applications that invoke ``critical balance'',
wave and nonlinear time scales are often set equal to one another, as a 
reflection of this balance.
In some related approaches,  
wave activity is presumed to be the dominant feature, and nonlinear 
effects are computed using 
perturbation theory.
In formal weak turbulence theory,
the leading order behavior is that of a wave, 
and nonlinearities that give rise to turbulence occur as 
next order corrections. 
Variations of this idea, often less formally implemented, 
include the ``wave turbulence" approach
~\cite{Cranmer2003ApJ, Chandran2005PRL, Chandran2008PRL, Howes2014Arxiv, Klein2012ApJ} 
that models finite amplitude 
plasma turbulence by approximating turbulent fluctuations as superposition 
of randomly-phased linear wave modes. 
This ``quasilinear premise'' performs well 
in the case of weak turbulence, 
and it is argued to 
remain somewhat valid even in strong turbulence (as in the case of {\it 
Goldreich and Sridhar}'s critical balance theory.)

For linear waves, the dynamics of a specified Fourier mode
is expected to be 
that of a stochastic oscillator rather than of a simple 
harmonic oscillator with a characteristic frequency, e.g. as pointed out by Matthaeus \textit{et al.}~\cite{Matthaeus2014ApJ}. 
For Alfv\'enic turbulence, TenBarge \textit{et al.}~\cite{TenBarge:CPC2014} 
attempted to model the effect of large scale modes, greater than 
the simulation system size,
by coupling individual Fourier modes  
$\mathbf{z^{\pm}}({\bf k})$ 
to an oscillating Langevin antenna. Dmitruk \textit{et al.}~\cite{Dmitruk2001ApJ} 
constructed a phenomenological model based on reduced MHD (RMHD) to describe the 
driving of turbulence in the open line regions of the corona, incorporating both wave 
propagation effects, and models for strong nonlinearity. More complex turbulence models such as these may clearly benefit from improved simple schemes to drive them at large scales or even in the inertial range. This provides a practical motivation for development in the present paper of an elementary model designed to emulate the statistics of MHD turbulence. Clearly there are a wide range of models and applications in which the 
relative balance
between wave activity and stochastic behavior become very important. 
Besides these two disparate qualitative
behaviors, one might expect 
the relative influence of driving and dissipation 
to enter as well. 
Our motivation in developing the present model 
is to provide a tractable simple dynamical model that demonstrates
these effects quantitatively with some level of consistency with 
turbulence solutions obtained by much more demanding DNS approaches.

The outline of the paper is as follows: in Sec.~\ref{sec:langevin}, we build up the basic equations. In Sec.~\ref{sec:nlinwave}, we model the nonlinear dissipation and Alfv\'enic wave term. In Sec.~\ref{sec:norm}, we normalize the equations. In Sec.~\ref{sec:inertial_modes}, we enforce the Kolmogorov's symmetries for inertial scale modes. In Sec.~\ref{sec:large_modes}, we build in
some scaling rules for the large scale modes. In Sec.~\ref{sec:numerical}, 
we discuss the numerical details. Sec.~\ref{sec:results} contains our results. 
We discuss some of the previous similar schemes and conclude in  Sec.~\ref{sec:discussion}.

\section{\label{sec:langevin}Langevin and single-mode models}
The one dimensional Langevin equation describing the Brownian motion of a particle is given by,
\begin{eqnarray}\label{eq:langevin}
	\frac{d u}{dt} = -\alpha u + g(t).
\end{eqnarray}
Where $u$ is the velocity of the particle. A systematic linear drag term $-\alpha u$ represents the friction and a fluctuating part $g(t)$ models the random kicks experienced by the particle from the surrounding medium. A Gaussian white noise is usually used for $g(t)$~\cite{Uhlenbeck:PR1930,Chandrasekhar:RMP1943}.

The Langevin equation and its solutions have many interesting 
properties and applications. 
Of historical importance is that 
its solutions provide a useful description of Browninan motion
~\cite{Smoluchowski1916PZ, Einstein1905AdP, Uhlenbeck:PR1930}.
One may also readily show that 
its solutions are equivalent to solutions of a Fokker-Plack equation
~\cite{Chandrasekhar:RMP1943}, and as such it can be 
related to various problems in diffusion theory and in
approaches to equilibrium distributions.
The addition of an oscillatory degree of freedom is 
quite relevant to the present model,
and was introduced~\cite{Balescu1994PoP}
to describe charge particle collisional diffusion 
including gyration in a uniform large scale magnetic field.
  
Our specific goal is to model the dynamical behavior of a single
Fourier mode in the inertial range of MHD turbulence. 
As such the physical structure and appropriate normalization of the 
basic incompressible model will be relevant.
The equations of incompressible MHD can be expressed in terms of the 
Els\"asser variables, $\mathbf{z_{\pm}} = \mathbf{u} \pm \mathbf{b}/\sqrt{4\pi\rho}$ as: 
\begin{equation}
\frac{\partial \mathbf{z}^{\pm}}{\partial t} =  \pm \mathbf{v_A} \cdot \nabla \mathbf{z}^{\pm} - \mathbf{z}^{\mp} \cdot \nabla \mathbf{z}^{\pm} - \frac{\nabla P}{\rho_0} - \mu \nabla^2 \mathbf{z}^{\pm}
\end{equation}
Where $\mathbf{u}$ and $\mathbf{b}$ are velocity and magnetic field 
fluctuations respectively. $\mathbf{v_{A}}$ is the Alv\'en velocity defined as 
$\mathbf{v_{A}} = \mathbf{B_0}/\sqrt{4\pi\rho_0}$, where $\mathbf{B_0}$ is the 
mean magnetic field. $P$ is the total pressure (thermal$+$magnetic), $\rho_0$ is 
the constant density and $\mu$ is the viscosity, assumed to be equal to the 
resistivity. The 
quadratic 
non-linear term, 
$\mathbf{z}^{\mp} \cdot \nabla \mathbf{z}^{\pm}$ and pressure term, $\nabla P/\rho_0$ 
are responsible for transferring energy from large scales, into and through the  
inertial range.  The plasma also  responds 
to the linear term, $\mathbf{v_A} \cdot \nabla \mathbf{z}^{\pm}$,
which induces wave-like dynamics.
The last term,
$\mu \nabla^2 \mathbf{z}^{\pm}$ is responsible for dissipation, and for large Reynolds numbers, 
this term is 
negligible except at very small 
scales.
For 
description of 
inertial range scales ($\gg$ dissipation range scales),
the explicit dissipation makes no contribution, 
its influence is manifested in the 
continual transfer of energy towards the smaller scales.
In quasi-steady conditions the transfer 
due to eddy motions will balance the 
viscous effects at the dissipation 
scales,
and therefore 
it is natural to suppose that inertial range
eddies provide a kind of effective dissipation, 
or eddy viscosity. 

A simple model was introduced by 
Dmitruk \textit{et al.}~\cite{Dmitruk2002ApJ}
to approximate the average damping experienced by a
a typical mode in the inertial range,
due to the 
net effect of all smaller scale modes.
Writing the MHD equations for 
a typical or representative mode at a selected scale $\ell$, 
the model
proceeds to account for nonlinear effects in a way 
that involves only the  amplitude at scale $\ell$ in accord with 
Kolmogorov's scale locality hypothesis. The nonlinearity takes the 
form of a quadratic drag force that is consistent with a
Taylor-von K\'arm\'an decay law for MHD~\cite{Hossain1995PoP}, 
and is a close relative of an 
eddy viscosity. 
This approach will be incorporated below. 

\section{\label{sec:model_develop}Development of the model}

We begin with the assumption that the effects of nonlinearity
are 
principally {\it local in scale}. 
Then 
a single Fourier mode (velocity or magnetic) residing in the inertial range, 
experiences a random net input of energy from the larger scales,
while it dissipates 
by transferring energy to smaller scales. On average, 
the input and dissipation balance, 
so that there is no secular pile up of energy at any scale
(no buildup of spectral amplitude at any wave number). 
Therefore, in analogy with the Langevin equation, 
Eq.~(\ref{eq:langevin}) we write a model equation for the velocity 
(and magnetic) fluctuation at length scale $l$,
\begin{eqnarray}\label{eq:modified_langevin}
	\frac{\partial u_{l}}{\partial t} = -\alpha u_{l} + g_{l}(t).
\end{eqnarray}

Here, we represent both velocity and magnetic fluctuation at length scale $l$ by $u_l$ for convenience. 
Note that unlike Eq.~(\ref{eq:langevin}), the fluctuating part $g_{l}(t)$ in Eq.~(\ref{eq:modified_langevin}) 
is not Gaussian white noise, but a chaotic force changing on a typical time scale $t_c$. 
(See Sec.~\ref{sec:numerical} for more discussion.) 
The amplitude of $g_{l}(t)$ as well as the time scale $t_c$ are expected to change 
with length scale $l$. In steady state magnetohydrodynamic (MHD) turbulence, 
usually a forcing function acts at some large scale $L$ which excites the nearby modes. 
A mode deep in the inertial range does not feel the same driving force. Rather, the influence of
 the force is cascaded by the larger scale couplings, through many intermediate modes,  
down to the scale of interest $l$. Here, we 
represent the driving felt at length scale $l$ by the term $g_{l}(t)$. 

We remark that Eq.~(\ref{eq:modified_langevin}), which remains formally a 
Langevin equation,
was also adopted by Beck~\cite{Beck:PRE1994}
for describing (or, as here, emulating)
properties of turbulence.
In contrast to the present case, 
Beck's Langevin model (Eq.~(\ref{eq:modified_langevin})) is written for
the evolution of an inertial 
range velocity {\it increment},
and the focus is on constructing a model 
that can 
reproduce scaling of higher order 
structure functions in the inertial range. 
Scaling (i.e., intermittency)
is introduced in the 
model through use of a forcing function 
based on a logistic map scheme.
As shown presently, 
our approach is quite 
different. 

\subsection{Modeling of nonlinearity and Alfv\'en waves}\label{sec:nlinwave}

To modify the classical Langevin equation, Eq.~(\ref{eq:langevin}), we start with the dissipative term $-\alpha u_{l}$. From von K\'arm\'an similarity hypothesis, extended to MHD, we expect
that~\cite{Karman1938PRSL, Hossain1995PoP},
\begin{eqnarray}\label{eq:vk}
	\frac{d u^2_l}{dt} \propto \frac{u^3_l}{l}.
\end{eqnarray}
Accordingly, 
we modify the linear drag term in Eq.~(\ref{eq:modified_langevin}) 
so that it becomes nonlinear, as,
\begin{eqnarray}
	\alpha u_l \rightarrow \frac{C}{l} |u_l|u_l.
\end{eqnarray}
and the model equation Eq.~(\ref{eq:langevin}) becomes,

\begin{eqnarray}\label{eq:model_vk}
	\frac{\partial u_{l}}{\partial t} = -\frac{C}{l} |u_{l}| u_{l} + g_{l}(t).
\end{eqnarray}
In this equation, if we set the forcing term to zero, 
the energy goes as,
\begin{eqnarray}
	\frac{\partial |u_{l}|^2}{\partial t} \sim - \frac{C}{l} |u_{l}|^3.
\end{eqnarray}

The constant $C$ appearing in Eq.~(\ref{eq:model_inertial}) and ~(\ref{eq:model_large}), is expected to be close to Taylor-von K\'arm\'an decay constant, extended to MHD~\cite{Hossain1995PoP, Wan2012JFM}. The value of $C$ has been measured to be around $0.5$ for hydrodynamic turbulence~\cite{Pearson2002PoF}. In MHD and kinetic plasmas, recent studies show hints of a universal decay rate at large Reynolds number, with a decay constant close to unity~\cite{Wu2013PRL,Parashar:ApJL2015}. We use a value of $C = 0.5$ in this study.

MHD turbulence also supports Alfv\'en waves,
most easily described by introducing a uniform mean magnetic field $\mathbf{B_0}$.
When present, the 
Alfv\'en wave frequencies are 
$\omega \sim \mathbf{k}\cdot\mathbf{V_A} \sim k B_0 \cos \theta $. 
To accommodate the physics of these incompressible 
MHD waves in the present model, 
we introduce a new term to the model equation Eq.~(\ref{eq:model_vk}),
which becomes
\begin{eqnarray}\label{eq:model_vk_wave}
	\frac{\partial u_{l}}{\partial t} = i \omega u_{l} -\frac{C}{l} |u_{l}| u_{l} + g_{l}(t).
\end{eqnarray}
Here $i=\sqrt{-1}$, and 
since $l \sim 1/k$, 
we have $\omega \sim B_0 \cos \theta/l$. 
Here $B_0$ is the characteristic magnetic field strength in Alfv\'en speed units. 
Note that the dependent 
variable $u_l$ in Eq.~(\ref{eq:model_vk}) 
has become complex, 
and represents a typical Fourier amplitude, 
and not simply a typical speed at scale $\ell$.
This is commensurate with the interpretation
of a typical Fourier mode in Dmitruk \textit{et al.}
\cite{Dmitruk2001ApJ}

Using this interpretation and our  
notation, $|u_{l}|^2$ can be equated to 
(twice) the energy per unit mass in fluctuations 
at a scale $l$ in turbulence. 
To make this correspondence we 
view $|u_l|$ as a contribution to 
a full (omnidirectional) 
spectrum $E(k)$ that varies over wavenumber $k$.
By construction, with $k\to 1/l$
\begin{eqnarray}
	&|u_l|^2 &= k E(k),\nonumber\\
	\mathrm{or,}&\quad E(k) &= l {|\hat u_l|^2}
\label{eq:u(k)_u_k}
\end{eqnarray} 
and therefore we can directly obtain a
spectrum from $u_l$ by varying $l$.

\subsection{\label{sec:norm}Normalization}

In order to compare the results from our model 
equation with other systems like DNS, 
we normalize our equations in a 
physically revealing way.
In the process, we shall also specify 
the characteristic timescale
(correlation timescale) 
that associated 
with $g_{l}(t)$. The following treatment will also enable us to specify an appropriate scale dependence of $g_l$, i.e., $g_{l}(t) = g(l,t)$.

To normalize  Eq.~(\ref{eq:model_vk_wave}), 
we select a large scale length $\lambda$, 
associated with which, we have a typical velocity (or magnetic) fluctuation 
$u_{0\lambda}$. This establishes 
the characteristic 
time scale $\tau_{0\lambda}$ 
through 
\begin{eqnarray}\label{eq:large_eddy}
	\tau_{0\lambda} = \frac{\lambda}{u_{0\lambda}}.
\end{eqnarray}
The scale $\lambda$ will represent the energy containing scale,
and so $\tau_{0\lambda}$ is the system eddy turnover time. 
Similarly, at another 
length scale of interest $l$, where for the inertial 
range, $l<\lambda$,
there is typical velocity 
$u_{0l}$ and an implied 
time scale $\tau_{0l}$ such that
\begin{eqnarray}\label{eq:inertial_eddy}
u_{0l} = \frac{l}{\tau_{0l}}.
\end{eqnarray}
We would like to emphasize the difference between 
$u_{l}$ and $u_{0l}$ or $\tau_{l}$ and $\tau_{0l}$. 
$u_{l}$ and $\tau_{l}$ are the instantaneous velocity fluctuation and time scale at length scale $l$ 
while $u_{0l}$ and $\tau_{0l}$ are the \emph{typical} velocity and time scale at same length scale $l$. 
With this, we normalize different quantities as follows,
\begin{eqnarray}\label{eq:nondim}
	u_{l} = u_{0\lambda} \widetilde{u_{l}},& \quad t = \tau_{0\lambda} \widetilde{t},& \quad l = \lambda \widetilde{l}.
\end{eqnarray}
Further, we write $g_l$ as,
\begin{eqnarray}\label{eq:modified_g}
	g_l = \bigg( \frac{u_{0l}}{\tau_{0l}} \bigg) \widetilde{g},
\end{eqnarray}
where $\widetilde{g}$ is a dimensionless order-one random function of time.
The quantities with $\quad \widetilde{} \quad$ in the overhead 
are all dimensionless, with $\widetilde{u_{l}}$ the dependent variable, 
$\widetilde{t}$ the independent variable, and 
$\widetilde{l}$ a fixed constant during time integration. 

Plugging Eq.~(\ref{eq:nondim}), ~(\ref{eq:modified_g}) into Eq.~(\ref{eq:model_vk_wave}), 
\begin{eqnarray}
	\frac{u_{0\lambda}}{\tau_{0\lambda}}\frac{\partial \widetilde{u}_{l}}{\partial \widetilde{t}} =&& i \omega u_{0\lambda} \widetilde{u_{l}} - i \omega u_{0\lambda} \widetilde{u}_{l}\nonumber\\
	 &&- \frac{C}{\widetilde{l}\lambda} u^2_{0\lambda} |\widetilde{u}_{l}|\widetilde{u}_{l} + \frac{u_{0l}}{\tau_{0l}}\widetilde{g},\\	 
	\frac{\partial \widetilde{u}_{l}}{\partial\widetilde{t}} =&& i (\omega \tau_{0\lambda}) \widetilde{u}_{l} - \frac{C}{\widetilde{l}}|\widetilde{u}_{l}|\widetilde{u}_{l} + \bigg( \frac{u_{0l} \tau_{0\lambda}}{\tau_{0l} u_{0\lambda}} \bigg) \widetilde{g}.
\end{eqnarray}

Now, we note that the wave time scale  can also be written in eddy turnover time units as
\begin{eqnarray}
	\omega \tau_{0\lambda} &=& k_{\parallel} B_{0} \tau_{0\lambda}\nonumber\\
    &=& (k_{\parallel}\lambda) \bigg( \frac{B_{0}}{u_{0\lambda}} \bigg)\nonumber \\
    &=& \widetilde{k}_{\parallel} \widetilde{B}_{0}\nonumber \\
    &=& \widetilde{\omega}\nonumber.
\end{eqnarray}

Therefore, we have,
\begin{eqnarray}\label{eq:nondim_eq}
	\frac{\partial \widetilde{u}_{l}}{\partial\widetilde{t}} =&& i \widetilde{\omega} \widetilde{u}_{l} - \frac{C}{\widetilde{l}}|\widetilde{u}_{l}|\widetilde{u}_{l} + \bigg( \frac{u_{0l} \tau_{0\lambda}}{\tau_{0l} u_{0\lambda}} \bigg) \widetilde{g}.
\end{eqnarray}

\subsection{\label{sec:inertial_modes}Inertial range modes}
In this section we aim to recover the Kolmogorov scaling $E(k) \sim k^{-5/3}$ for the inertial range in a systematic way. For $l$ in the inertial range,
following Kolmogorov~\cite{Kolmogrorov1941a}, 
we enforce the following symmetries in Eq.~(\ref{eq:nondim_eq}),
\begin{eqnarray}
	\bigg( \frac{\tau_{0l}}{\tau_{0\lambda}} \bigg) &=& \bigg( \frac{l}{\lambda} \bigg)^{2/3},
\label{eq:kolmo_symm_t}\\
	\bigg( \frac{u_{0l}}{u_{0\lambda}} \bigg) &=& \bigg( \frac{l}{\lambda} \bigg)^{1/3}.\label{eq:kolmo_symm_u}
\end{eqnarray}
So, Eq.~(\ref{eq:nondim_eq}) becomes
\begin{eqnarray}\label{eq:model_inertial}
	\frac{\partial \widetilde{u}_{l}}{\partial\widetilde{t}} =&& i \widetilde{\omega} \widetilde{u}_{l} - \frac{C}{\widetilde{l}}|\widetilde{u}_{l}|\widetilde{u}_{l} +  \widetilde{l}^{-1/3} \widetilde{g}.
\end{eqnarray}	
This is the equation we propose for emulation of 
inertial range modes. 

From Eq.~(\ref{eq:model_inertial}) we can write,
\begin{eqnarray}\label{eq:energy_inertial}
	\quad \frac{\partial |\widetilde{u}_{l}|^2}{\partial\widetilde{t}} =&&  -2\frac{C}{\widetilde{l}}|\widetilde{u}_{l}|^3 +  \widetilde{l}^{-1/3} (\widetilde{g}\widetilde{u}^{*}_{l} + \widetilde{g}^{*}\widetilde{u}_{l})	
\end{eqnarray}
In the steady state, $\frac{\partial |\widetilde{u}_{l}|^2}{\partial\widetilde{t}} = 0$ and $|\widetilde{g}| \sim 1$. So,
\begin{eqnarray}\label{u2_ss}
	\frac{C}{\widetilde{l}}|\widetilde{u}_l|^3 &\sim& \widetilde{l}^{-1/3} |\widetilde{u}_{l}|,\\
	|\widetilde{u}_l|^2 &\sim& \widetilde{l}^{2/3},
\end{eqnarray}
which is consistent with Kolmogorov's theory (See Eq.~(\ref{eq:kolmo_symm_u})). 

The random term $\widetilde{g}$ fluctuates randomly in time and it has a correlation time $t_c$ associated with it. For a particular mode to be driven efficiently, one expects that the correlation time, $t_c$ has to be of the same order of the local turbulence time scale $(\tau_l)$(See~\cite{Dmitruk:APJ2003, Parashar:PoP2011, TenBarge:CPC2014}). Therefore we equate the correlation time with the local turbulence time scale,
\begin{eqnarray}\label{eq:tc_inertial}
\widetilde{t}_c = \frac{\tau_{0l}}{\tau_{0\lambda}}.
\end{eqnarray}

\subsection{\label{sec:large_modes}Large scale modes}
Large scale modes do not exhibit universal scaling in three dimensional MHD turbulence. 
Still there are some common characteristics observed in large scale modes in MHD turbulence. One such feature is presence of $1/f$ or ``flicker noise'' in the lowest wavenumber mode. Presence of low frequency $1/f$ signal in turbulent systems was investigated in some 
detail by Dmitruk and Matthaeus~\cite{Dmitruk:PRE2007}, and Dmitruk \textit{et al.}~\cite{Dmitruk:PRE2011}. 
A heuristic reasoning was proposed in~\cite{Dmitruk:PRE2007} and ~\cite{Dmitruk:PRE2011} to explain the occurrence of low frequency $1/f$ noise in the lowest wavenumber mode. 
The lowest wavenumber mode, 
say $k=1$, interacts with other modes via triadic interaction of the 
(schematic) form
\begin{eqnarray}
	\frac{\partial b(k=1)}{\partial t} \sim  i \sum_{p+q=1} u(q) b(p)\label{eq:triad_int},
\end{eqnarray}
Where $b(k=1)$ represents the first mode, and $u(p), u(q)$ are generic Fourier mode amplitudes. 
If the interaction is nonlocal, $p, q \gg k=1, p \sim q$, the time scale from the right hand side of Eq.~(\ref{eq:triad_int}) is $[ u(q)b(q)/b(k=1) ]^{-1}$, which is much longer than the local eddy turn over time scale since $u(q), b(q) \ll u(k=1), b(k=1)$. It is worth mentioning here that $1/f$ signal has also been observed in other turbulent plasma systems like interplanetary magnetic field~\cite{Matthaeus1986PRL}, solar corona~\cite{Bemporad2008ApJL} and photosphere~\cite{Matthaeus2007ApJL} although the source of $1/f$ signal in these systems may be (at least partially) different from the above heuristic reasoning.

One might 
anticipate the  emergence of low frequency $1/f$ noise in a
Langevin model such as Eq.~(\ref{eq:model_inertial}), 
if, for the case in question, 
one imposes 
a time scale ($t_c$) of the stochastic forcing $\widetilde{g}$, 
that is much greater than the local eddy turn over time $(l/|u_{l}|)$. 
We demonstrate this in the following way. We postulate 
(in an {\it ad hoc} fashion)
that for the large 
scale modes with $l>\lambda$, 
the 
time scale go as square of length scale. 
So we have,
\begin{eqnarray}
\bigg( \frac{\tau_{0l}}{\tau_{0\lambda}} \bigg) &=& \bigg( \frac{l}{\lambda} \bigg)^{2}.\label{eq:large_symm_t}
\end{eqnarray}
We further assume that in this range of scales, the 
characteristic speeds are simply 
proportional to the respective
characteristic lengths, i.e., 
\begin{eqnarray}
\bigg( \frac{u_{0l}}{u_{0\lambda}} \bigg) &=& \bigg( \frac{l}{\lambda} \bigg).\label{eq:large_symm_u}
\end{eqnarray}
So, for large scale modes, Eq.~(\ref{eq:nondim_eq}) becomes
\begin{eqnarray}\label{eq:model_large}
\frac{\partial \widetilde{u}_{l}}{\partial\widetilde{t}} =&& i \widetilde{\omega} \widetilde{u}_{l} - \frac{C}{\widetilde{l}}|\widetilde{u}_{l}|\widetilde{u}_{l} +  \widetilde{l}^{-1} \widetilde{g}.
\end{eqnarray}
Also we equate the correlation time of the forcing, $t_c$ with the time scale in Eq.~(\ref{eq:large_symm_t}),
\begin{eqnarray}
\widetilde{t}_c = \frac{\tau_{0l}}{\tau_{0\lambda}} = \bigg( \frac{l}{\lambda} \bigg)^{2}\label{eq:tc_large}.
\end{eqnarray}

We choose $\lambda = $ the length scale which separates the large scale modes from the inertial modes. 
So, for $l < \lambda$, 
we use the Eq.~(\ref{eq:model_inertial}) and for 
$l > \lambda$, we use Eq.~(\ref{eq:model_large}). 
Note that in this notation, the non-dimensional wavenumber $\widetilde{k} = 1/\widetilde{l} = \lambda/l$ can 
be greater (inertial scales), equal, or less (large scales) than one. 
From now on, for convenience, we 
will use the variables without the $\quad \widetilde{} \quad$ in the overhead. 

\begin{figure}
\begin{center}
	\includegraphics[scale=1.0]{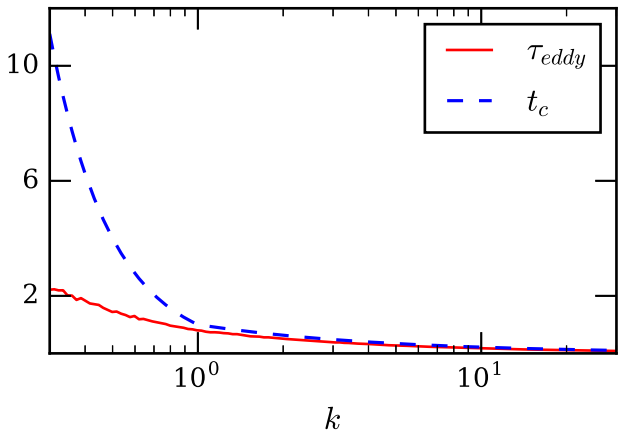}
	\caption{Plot of correlation time scale of the stochastic forcing, $t_c$ along with calculated eddy turn over time $\tau_{eddy} \sim l/|u_{l}|$, obtained from the model. For the $\tau_{eddy}$ calculation, we use time averaged value of $|u_{l}|$. All the variables are dimensionless as discussed in the text.}
	\label{fig:tscale_compare}
\end{center}
\end{figure}	

\section{\label{sec:numerical}Numerical Details}
For numerical implementation, Eq.~(\ref{eq:model_inertial}) and ~(\ref{eq:model_large}) 
are advanced in time 
using the fourth-order Runge-Kutta (RK4) scheme. 
The nondimensional wavenumber $k$ varies logarithmically
from $0.3$ to $30$. 
The total number of points in wavenumber space is $128$. 
The variables used in Eq.~(\ref{eq:model_inertial}) have a real 
and an imaginary component. Numerically, we determine the random forcing term 
$g$ as,
\begin{eqnarray}\label{eq:g_num}
	g(t) = m g(t-\Delta t) + \sqrt{(1-m^2)}  r(t).
\end{eqnarray}
The value of $g$ at $(n+1)$th step i.e. at time $t = (n+1)\Delta t$, is generated 
from the previous value (at $n$th step) of $g$ and a random number, $r$. 
The constant $m$ is a memory constant
that 
relates how the sequence ``remembers'' the past values. 
The random number r satisfies the following properties,
\begin{eqnarray}
\langle r \rangle &=& 0,\label{eq:r_assump1}\\
\langle r(t) r(t^{\prime}) \rangle &=& \delta(t-t^{\prime}).\label{eq:r_assump2}
\end{eqnarray} 
These are satisfied by a normalized Gaussian random number generator. 
With these properties one can derive the correlation function for 
$g$ (See ~\cite{BreechPhDThesis}),
\begin{eqnarray}\label{eq:g_corr}
	G_c(\tau) &=& \langle g(t)g(t + \tau) \rangle \nonumber\\
			  &=& e^{-\tau/t_c},\\
	\mathrm{and,}\quad t_c &=& \frac{\Delta t}{1-m}.
\end{eqnarray}
These results are valid in the appropriate 
limit, with, e.g., 
$\Delta t \to 0$, and $n\to \infty$, such that $\tau=n\Delta t$ remains constant, $t$ 
similarly defined, $m \to 1$ with $t_c$ constant,
and 
Eqs.~(\ref{eq:r_assump1}), (\ref{eq:r_assump2}) enforced.
Also note
that 
the real and imaginary part of the random term 
$\widetilde{g}$ in Eq.~(\ref{eq:model_inertial}) are independent of each other. 
That is, at each time step the real and imaginary components 
are determined from 
Eq.~(\ref{eq:g_num}) with same memory 
function $m$, but employing different 
random numbers.

We choose a time step $\Delta t = 10^{-3}$, and we run the simulation for $10^{9}$ time steps i.e. $10^{3}$ units of nondimensionalized time. 

For calculating frequency spectra, we use an ensemble average of ten independent time series with random initial conditions. This helps eliminate unwanted noise in the Fourier spectrum.

\section{\label{sec:results}Results}
\subsection{\label{sec:isotropic}Isotropic Case}

Following the numerical method described in the preceding 
section, we obtain solutions for 
with varying length scale $l$ (reciprocal of the wavenumber $k$.  
For the results shown, $k\lambda$ ranges from $0.3$ to $30$.
In this section,
we consider the case $\omega = 0$, which corresponds to the situation of globally 
isotropic MHD with no mean magnetic field, $\mathbf{B_0}=0$. 

As a first step, we
examine the relationship between the imposed 
correlation time scale $t_c$ of the forcing $g$,
with the local eddy turn over time $(\tau_{eddy} \sim l/|u_{l}|)$, since we 
control the former 
while the latter is determined from the solution.
Fig.~\ref{fig:tscale_compare} shows this comparison.
For  the $\tau_{eddy}$ calculation, we use time averaged value of $|u_{l}|$. 
At smaller scales 
associated with the inertial range, 
the characteristic local nonlinear time
was chosen to follow a Kolmogorov 
scaling (Eq.~\ref{eq:kolmo_symm_t})
and the the forcing correlation time and 
computed eddy turnover time are found to be
almost equal. 
However, 
for the long wavelength modes with $l>\lambda$
the imposed scaling was modified 
(Eq. \ref{eq:large_symm_t}) to favor longer timescales, 
and we find from the solutions that $t_c \gg \tau_{eddy}$ for the lowest wavenumber
$k\lambda =0.3$. 
This establishes the long wavelength conditions that 
have been argued \cite{Dmitruk:PRE2007} to favor
generation of a $1/f$ signal in the {\it frequency} spectra, an issue we address below.

The distribution of energy over wavenumber 
(inverse length scale)
is a central issue in turbulence,
and our simple model is designed to 
reproduce this basic result. 
In contrast to hydrodynamics, 
the energy spectra in MHD is perennially debated and there have been a number of 
proposed phenomenological models (See~\cite{Verma2004PR} for a review). 
As an example, 
using a renormalization group technique, Verma~\cite{Verma:PoP1999} showed that
isotropic MHD should follow a $-5/3$ energy spectrum in the inertial range. 
Solar wind observation and DNS results also favor Kolmogorov like spectra. 
In the present model
wave number $k$ and length scale $l$ are 
related simply by $k \sim 1/l$ 
and the energy spectrum 
corresponds to $E(k) \to l\langle u_l{^2}\rangle $.

\begin{figure}
\begin{center}
\includegraphics[scale=1.0]{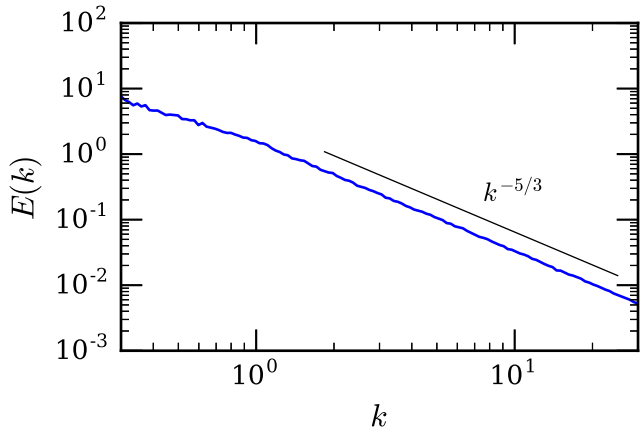}
\caption{Energy spectra $E(k)$ obtained from the model with $\omega = 0$. 
Kolmogorov scaling, $\sim k^{-5/3}$ scaling is shown as reference.}
\label{fig:iso_spectrum}
\end{center}
\end{figure}

Figure~\ref{fig:iso_spectrum} shows the spectrum
$E(k)$ as function of wavenumber $k$ using our model. 
Kolmogorov $k^{-5/3}$ scaling is evident at large $k$. 
At small wavenumbers, $k <1$, 
the spectrum scales approximately as $k^{-1}$; 
this is a consequence of the particular choice 
of long wavelength time scales Eq. \ref{eq:large_symm_t} 
and 
should not be viewed as of essential import.
For each $k$, we have time averaged $|u_l|^2$ to 
obtain $E(k)$. 

The scale dependent two-time correlation function 
for a full Fourier representation of the Elas\"asser variables
is defined as,
\begin{eqnarray}\label{eq:2pt_autocorr}
	R(\mathbf{k},\tau) = \frac{\langle \mathbf{z}^{\pm}(\mathbf{k},t) \cdot \mathbf{z}^{\pm *}(\mathbf{k},t+\tau) + \mathrm{c.c.} \rangle}{\langle |\mathbf{z}^{\pm}(\mathbf{k})|^2 \rangle}.
\end{eqnarray}
Where $\langle \cdots \rangle$ represents average over all time $t$, $\mathrm{c.c.}$ stands for complex conjugate, $\mathbf{k}$ is a specified wave vector, $\tau$ is the time lag and $\mathbf{z}^{\pm}(\mathbf{k},t)$ is the Fourier amplitude of $\mathbf{z}^{\pm}$ at wave vector $\mathbf{k}$ at time t. 
Scale dependent time correlations are well recognized as 
essential elements of turbulence theory \cite{Orszag1972PRL}
including closures \cite{McComb1990Book}. 

\begin{figure}
\begin{center}
\includegraphics[scale=1.0]{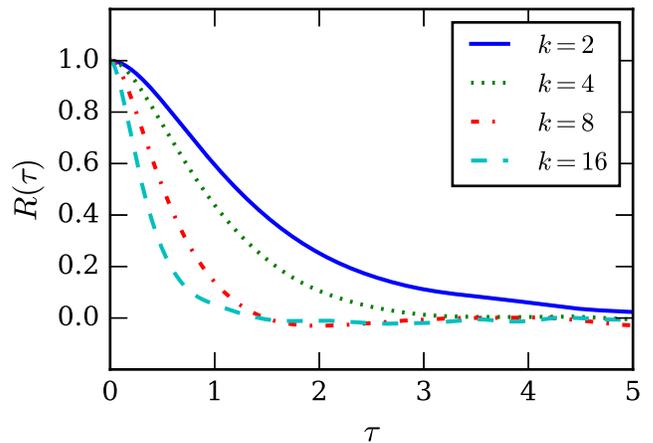}
\caption{Correlation function $R(\tau)$ as function of the time lag $\tau$ 
for different wavenumbers $k$.}
\label{fig:iso_R}
\end{center}
\end{figure}

\begin{figure}
\begin{center}
\includegraphics[scale=1.0]{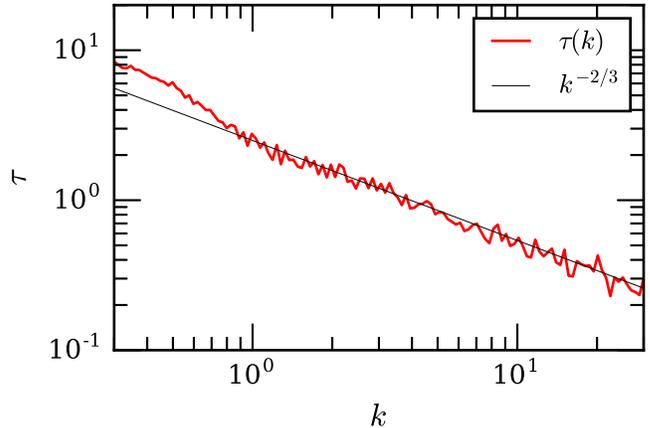}
\caption{Correlation time $\tau$ as a function of wavenumber 
$k$ from the model equation. For comparison, nonlinear time, $\sim k^{-2/3}$ is shown.}
\label{fig:iso_tau-k}
\end{center}
\end{figure}

For our model equation, we calculate the two-time autocorrelation function as,
\begin{eqnarray}\label{eq:2pt_autocorr_model}
R(k,\tau) = \frac{\langle u_k(t) u_k^{*}(t+\tau) + \mathrm{c.c.} \rangle}{\langle |u_k|^2 \rangle}.
\end{eqnarray}
Where, it is understood that $k = 1/l$ and $u_k = u_l$. 
Numerically, for computing the autocorrelation function, 
we use the Blackman-Tukey technique~\cite{Matthaeus:JGR1982},
which gives a consistent 
approximation 
based on the formula
\begin{eqnarray}\label{eq:Blackman-Tukey}
\langle |u_k|^2 \rangle R(n) = \frac{1}{M-n} \sum_{P=1}^{M-n} && \left[ u_k(P) u_k^{*}(P+n)+ \mathrm{c.c.}\right]\\
&& n = 0, 1, 2, \cdots N.\nonumber
\end{eqnarray}
Here, N is the maximum lag for which $R$ is calculated and M is the total number of data points. Data points are separated by time step $\Delta t$. 
We computed the two-time correlation function 
obtained from solutions of the model equation for varying scale size $l$. 
The results are shown in Figure~\ref{fig:iso_R}.
Previously, Servidio {\em et al.}~\cite{Servidio:EPL2011} and Lugones {\em et al.}~\cite{Lugones:PoP2016} calculated the two-point (scale dependent) 
temporal autocorrelation functions for isotropic MHD turbulence. 
These studies found that the correlation function $R(k,\tau)$ approaches 
zero with different rates for different wave vector $k$. The higher $k$s 
decay at faster rate compared to lower $k$s. Similar behavior can be 
seen from plots in Figure~\ref{fig:iso_R}. Once the calculation of 
$R(\tau)$ is done, we can also calculate the correlation time $\tau(k)$
as a 
function of effective wavenumber $k=1/l$, 
\begin{eqnarray}
\tau(k) = \int_{0}^{\infty} R(\tau) d\tau.\label{eq:tau_k}
\end{eqnarray}
In practice the upper limit of the integral extends to the 
final time in the data record. The plots of correlation time as 
function of $k$ is shown in Figure~\ref{fig:iso_tau-k}.
Servidio {\em et al.}~\cite{Servidio:EPL2011} used a best fit of 
six data points and found that the scaling of $\tau(k)$ with $k$ is 
closer to the sweeping time $\tau_{sw}$, which goes as $k^{-1}$, than 
the local nonlinear time $\tau_{nl}$, which scales like $k^{-2/3}$. 
Lugones {\em et al.}~\cite{Lugones:PoP2016} used more dense data points 
in $k$ space and showed that, for isotropic case, the correlation time 
$\tau(k)$ is close to the sweeping time $\tau_{sw} \sim k^{-1}$ for 
small $k$ and it is close to the nonlinear time $\tau_{nl} \sim k^{-2/3}$ at 
large $k$ values. 
In Figure~\ref{fig:iso_tau-k}, we see that in the inertial range, the correlation time $\tau(k)$ scales like $k^{-2/3}$, rather than $k^{-1}$, as observed in isotropic MHD simulations~\cite{Lugones:PoP2016}. This indicates that our model does not capture the sweeping effects properly. We discuss this issue more elaborately later in the paper.

\begin{figure}
\begin{center}
\includegraphics[scale=0.65]{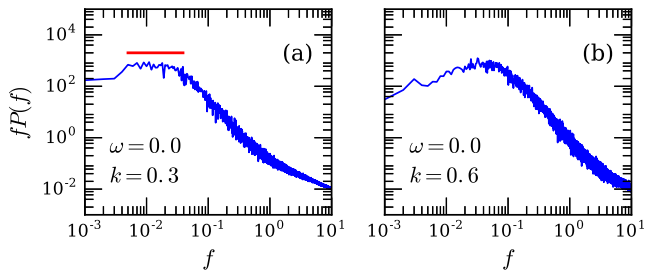}
\caption{Ensemble averaged compensated frequency power spectrum. A horizontal bar in panel (a) indicates the region of approximate $1/f$ spectral behavior. No such region is prominent in panel (b).}
\label{fig:iso_1overf}
\end{center}
\end{figure}

Complementary views are provided by alternative
analyses of the space-time correlations.   
For example scale the dependent correlation functions, 
as in Eq. \ref{eq:2pt_autocorr}, or more specific to our model, 
Eq. \ref{eq:2pt_autocorr_model}, may also be Fourier analyzed 
in time. This leads to scale (or wave vector) dependent 
frequency spectra, examples of which are shown 
for model results in Figure~\ref{fig:iso_1overf}.
In this figure, 
$P(f)$ is the square absolute value of the complex 
Fast Fourier Transform (F.F.T.) of the time series $u_k(t)$
\begin{eqnarray}
	\hat{u}_k(f) &=& \mathrm{F.F.T.}[u_k(t)],\\
	P(f) &=& |\hat{u}_k(f)|^2.
\end{eqnarray}

In three dimensional, isotropic MHD, a $1/f$ signal is observed in the frequency spectra, 
in the largest mode, while 
in three dimensional hydrodynamics (HD), $1/f$ noise is absent, or at least weaker than MHD ~\cite{Dmitruk:PRE2007}. In Figure~\ref{fig:iso_1overf} we show the ensemble averaged compensated frequency space spectrum for $k = 0.3$, which is the largest mode in our system, and for one of the large modes, $k = 0.6$. We have plotted the product of frequency by spectral power, $fP(f)$ to facilitate identification of $1/f$ power.
Both axes are in nondimensional units as discussed in Sec~\ref{sec:norm}.
For $k = 0.3$ mode, $1/f$ noise appears to be present for more than one decade while it is absent or at least less clear for $k = 0.6$ mode. 
This is a consequence of the time scales that we imposed on the large scale modes, 
given in Eq. \ref{eq:large_symm_t}.
However the result does demonstrate
the plausibility of the explanation for $1/f$ noise advanced 
by Dmitruk \textit{et al.}\cite{Dmitruk:PRE2007}, as further discussed below.

\begin{figure}
\begin{center}
\includegraphics[scale=1.0]{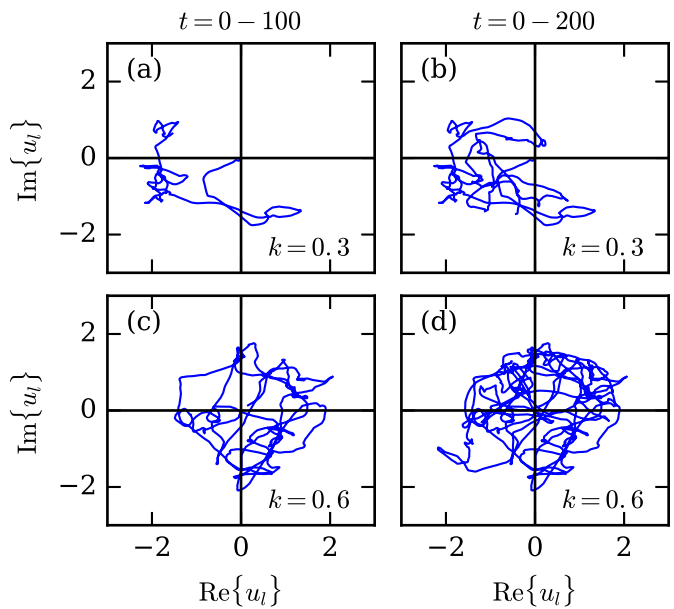}
\caption{Phase space plots of $k=0.3$ and $k=0.6$ modes with $\omega=0$ as obtained from the model equation. Panel (a) and (b) show the trajectory of $k=0.3$ mode at time $t=0-100$ and $t=0-200$ respectively. Panel (c) and (d) show the trajectory of $k=0.6$ mode at time $t=0-100$ and $t=0-200$ respectively.}
\label{fig:iso_phase}
\end{center}
\end{figure}

Dmitruk {\em et al.}~\cite{Dmitruk:PRE2011} argued that the emergence of $1/f$ fluctuations 
in the frequency spectra of the 
largest modes in MHD is related to the long range interaction between the 
largest mode and the small scale modes. The largest mode spends more times 
in clusters of phase space before jumping to a different region 
and thereby covers less area in the phase space in a given time interval, 
compared to the large scale modes. This phenomenon is a manifestation 
of broken ergodicity \cite{Shebalin:GAFD2013, Servidio:PRE2008a}
and is thought to be related to the appearance of $1/f$ noise. 
We analyze the phase space diagram for the $k = 0.3$ and $k = 0.6$ modes in 
Figure~\ref{fig:iso_phase}. The left and right panels in each set of figures show 
the evolution of the phase space trajectory in same intervals of times for the two modes. 
It is clear that in the same interval of time, the 
$k=0.3$ mode covers less area of the phase space compared to the $k = 0.6$ mode. The $k=0.6$ mode covers the phase space more uniformly compared to the $k=0.3$ mode which mostly remains limited to the left quadrants during the time considered.

\subsection{\label{sec:anisotropic}Anisotropic Case: Effect of the Imposed DC Magnetic Field}
We perform a set of simulations of the model equation with $\omega = 0.5, 1.0, 2.0$. This may be related to MHD turbulence in presence of a dc mean magnetic field $\mathbf{B_0}$. 
It is well known that turbulence in MHD with a dc field $\mathbf{B_0}$ 
has a strong tendency to become anisotropic, the cascade
producing gradients stronger in the two directions perpendicular to $\mathbf{B_0}$,
and weaker in the parallel direction \cite{Shebalin1983JPP, Oughton1994JFM}.
Physically this occurs because Alfv\'en wave-like couplings destructively 
interfere with correlations that establish nonlinear spectral transfer. 
This suppresses parallel spectral transfer but has no direct effect on 
perpendicular transfer. 

The present model is in effect one-dimensional, so it is not possible 
to fully represent the anisotropic dynamics anticipated for 
a full MHD representation. 
However from the developments leading to the model equation above, 
we can compare the frequency dependent wave term to the nonlinear term, 
for example in Eq. \ref{eq:nondim_eq}.
The ratio is $\frac{k_\parallel V_A}{u_l/l} =\cos{\theta} \frac{V_A}{u_l}$.
Consequently if we choose a scale $l$, a turbulence level and a
mean magnetic field strength, then the strength of the Alfv\'en 
wave effect is controlled only by the angle $\theta$ between 
${B}_0$ and ${\bf k}$. 

While we again emphasize that a simplified one 
dimensional model cannot capture the full physics 
of anisotropic MHD, 
by varying $\omega$ we can examine how changing the 
applied field strength (or, equivalently, 
the angle $\theta$) affects the solutions. 
 
There is another, more subtle way that 
wave activity may enter. 
Consider modes that have 
wave vector purely perpendicular to the direction of 
${\bf B}_0$, so that $k_{\parallel} = 0$.
One might naively expect that these $k_{\perp}$ modes (also called ``2D modes'')
would not experience wave signatures since their intrinsic
Alfv\'en frequency $\omega = k_{\parallel} B_0 = 0$.
However, through nonlinear interactions,
the 2D/$k_{\perp}$ modes are coupled to 
modes with nonzero 
$k_{\parallel}$ modes. 
Dmitruk and Matthaeus~\cite{Dmitruk:PoP2009} studied this 
phenomena in some detail, finding 
the presence of wave activity in 
certain $k_\parallel=0$ modes 
manifested by appearance of Alfv\'en frequency
peaks in the frequency spectrum of $k_{\perp}$ modes.
Howes and Nielson~\cite{Howes:POP2013} 
derived some elementary nonlinear couplings of this 
type by considering collision of Alv\'en waves 
in the weakly nonlinear limit. 
One concludes that the 
effective value of the Alfv\'en frequency $\omega$ for a mode 
with specified wave vector $k_{\perp}$ depends on 
which $k_{\parallel}$ mode interacts most dominantly. 
This may vary in different systems with different setups. 

In principle a Fourier mode with wave vector perpendicular 
to the mean magnetic field direction can feel a frequency value anywhere in the 
range $0 < \omega < \infty$. 
However, for direct wave couplings,
the increasing values of $\omega$ is qualitatively 
similar to the effect of increasing dc field while holding 
$k_{\parallel}$ constant or 
keeping the dc field constant while increasing $k_{\parallel}$, up to a maximum value 
of $k_\parallel = k = 1/l$. 
Indirect wave couplings may be qualitatively studied by 
judiciously selecting a nonzero $\omega$ for any chosen 
$k_{\perp}$ mode. 

\begin{figure}
\begin{center}
\includegraphics[scale=1.0]{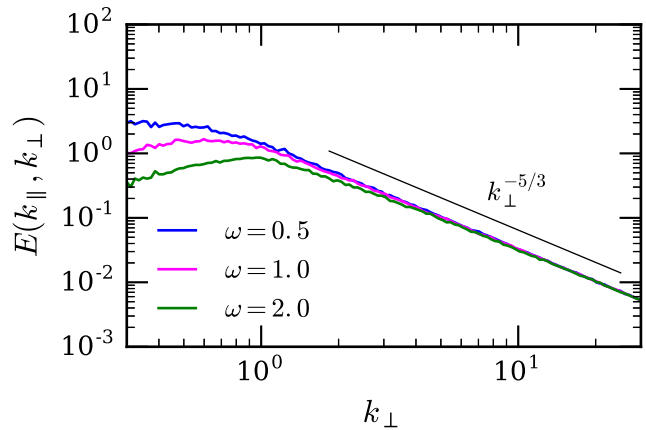}
\caption{Energy spectra $E(k_\parallel, k_{\perp})$ vs $k_\perp$
obtained from the model with $\omega = 0.5, 1, $ and $2$. Kolmogorov scaling, $\sim k_{\perp}^{-5/3}$ is shown for reference.}
\label{fig:aniso_spectrum}
\end{center}
\end{figure}

Figure~\ref{fig:aniso_spectrum} shows the energy spectrum plots obtained from the 
model for the three cases, $\omega = 0.5, 1, 2$. 
One may interpret this as an illustration of the 
anisotropic  spectrum $E(k_\parallel, k_\perp)$ plotted as functions of $k_\perp$,
for a series of three values of $k_\parallel$.
A spectral form $E(k_{\perp}) \sim k^{-5/3}_{\perp}$ 
is clearly present in the inertial range.

\begin{figure}
\begin{center}
\includegraphics[scale=1.0]{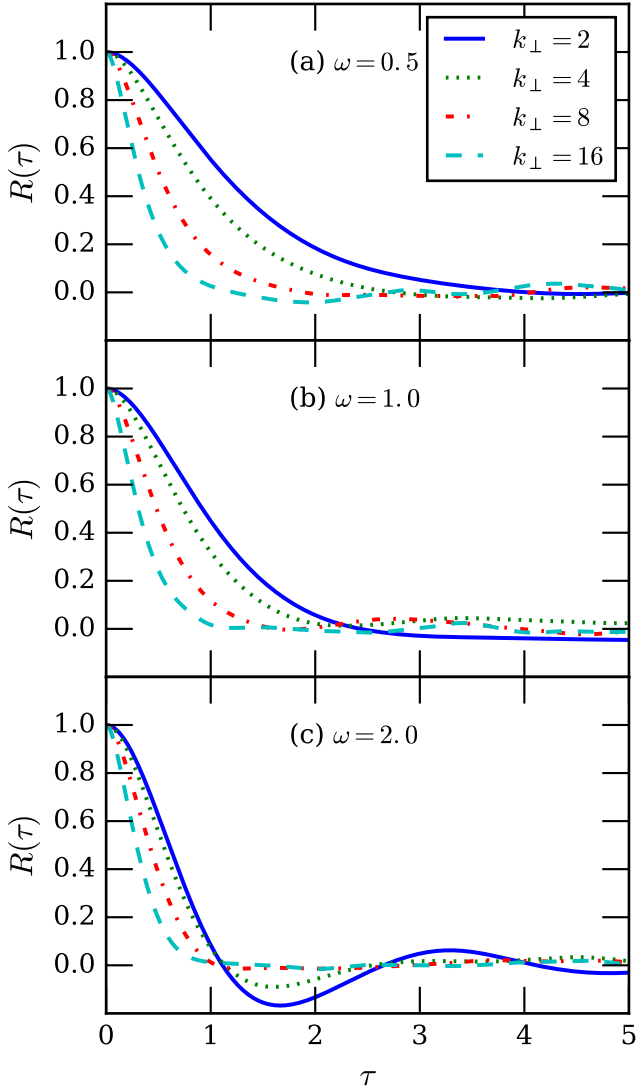}
\caption{Correlation function $R(\tau)$ as function of the time lag $\tau$ for different $k$ modes with nonzero $\omega$. In each panel, $\omega$ is constant; (a) $\omega = 0.5,$ (b) $\omega = 1,$ (c) $\omega = 2$.}
\label{fig:aniso_corrfn}
\end{center}
\end{figure}

Following the same procedure as mentioned for the $\omega = 0$ case, we calculate the two-time autocorrelation function from the model for different values of $\omega$. Results are shown in Figure~\ref{fig:aniso_corrfn}. Similar to the $\omega = 0$ case, the correlation function $R(k,\tau)$ drops faster for larger $k$, but the disparity becomes less prominent with larger $\omega$. This is consistent with the DNS results as found in ~\cite{Servidio:EPL2011,Lugones:PoP2016}.

\begin{figure}
\begin{center}
\includegraphics[scale=1.0]{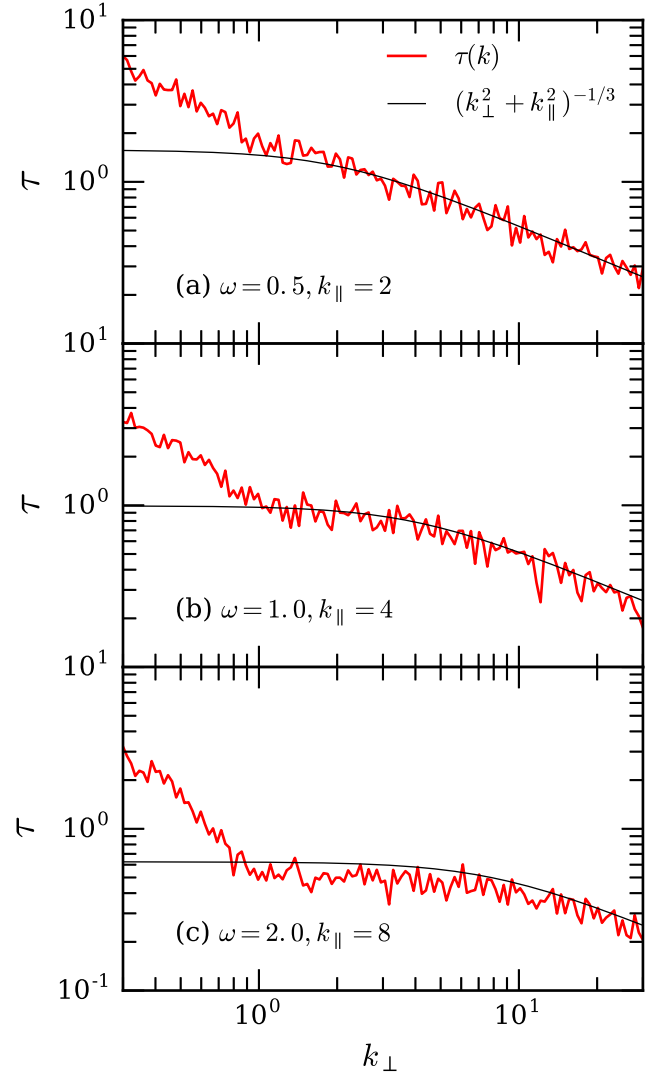}
\caption{Correlation time $\tau$ as a function of perpendicular wavenumber $k_{\perp}$ for different values of $\omega$; (a) $\omega = 0.5,$ (b) $\omega = 1,$ (c) $\omega = 2$. The dashed curves show a fitted nonlinear time, $\sim (k^2_{\perp}+k^2_{\parallel})^{-1/3}$ for comparison. (a) $k_{\parallel} = 2.0$, (b)  $k_{\parallel} = 4.0$, (c) $k_{\parallel} = 8.0$.}
\label{fig:aniso_tau-k}
\end{center}
\end{figure}

An interesting observation can be made from Figure~\ref{fig:aniso_tau-k}. In Figure~\ref{fig:aniso_tau-k} (a), (b), and (c) we have plotted the decorrelation time $\tau$ as function of wavenumber $k_{\perp}$, for $\omega = 0.5, 1.0$ and $2.0$ respectively, as obtained from the model equation. 
We fit the data with nonlinear time $\tau_{nl} \sim (k_{\perp} + k_{\parallel})^{-1/3}$, for $k_{\parallel}$ in the ratio of $1:2:4$ for the three cases $\omega = 0.5, 1.0, 2.0$. We found reasonably good fit for $k_{\parallel} = 2, 4,$ and $8$ for (a), (b), (c). The fitting curves have been scaled by multiplying a constant to align with the data. Like the $\omega=0$ case, we can compare Figure~\ref{fig:aniso_tau-k} with the findings of Lugones \textit{et al.}~\cite{Lugones:PoP2016} where it was found that, for the anisotropic case, the effect of Alfv\'en frequency enters the decorrelation time $\tau(k)$ in a way that $\tau(k)$ follows the sweeping time scale, $\tau_{sw}$ more closely than the nonlinear time $\tau_{nl}$. As mentioned before, since we do not consider the effects of sweeping eddies in the present model, we expect the decorrelation time to scale like local nonlinear time as found in Figure~\ref{fig:aniso_tau-k}. We come back to the discussion on sweeping effects in Sec.~\ref{sec:discussion}.

\begin{figure}
\begin{center}
\includegraphics[scale=0.65]{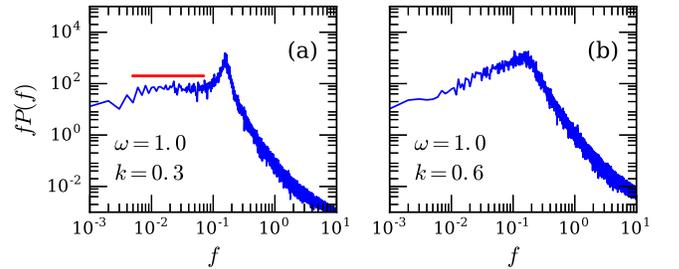}
\caption{Ensemble averaged compensated frequency power spectrum. A horizontal bar in panel (a) indicates the region of approximate $1/f$ spectral behavior. No such region is prominent in panel (b).}
\label{fig:1overf_1}
\end{center}
\end{figure}

Dmitruk and Matthaeus~\cite{Dmitruk:PRE2007} showed that $1/f$ 
signal appears at low frequencies in the frequency spectrum 
and the signal becomes stronger with increasing mean field. 
In Figure~\ref{fig:1overf_1} we show the compensated frequency spectrum obtained from the model for $\omega=1$. Again, a clear presence of $1/f$ power can be seen in the $k=0.3$ mode in the left panel while the $k=0.6$ mode show no such scaling. Also, comparing Figure~\ref{fig:iso_1overf}(a) and Figure~\ref{fig:1overf_1}(a), the range of $1/f$ power for the $\omega=1.0$ case is somewhat larger than the $\omega=0.0$ case. Therefore, it can be concluded that the $1/f$ signal becomes stronger with increasing $\omega$, as seen in MHD simulations~\cite{Dmitruk:PRE2007}.

\begin{figure}
	\begin{center}
		\includegraphics[scale=1.0]{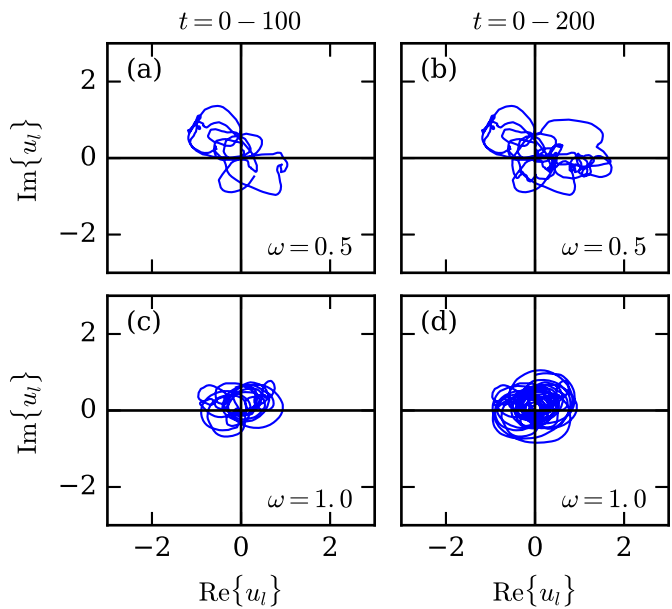}
		\caption{Phase space plot of the mode $k=0.3$ for $\omega=0.5$ and $\omega=1$. Panel (a) and (b) show the trajectory of the $k=0.3$ mode for $\omega=0.5$ at time $t=0-100$ and $t=0-200$ respectively. Panel (c) and (d) show the trajectory of the $k=0.3$ mode for $\omega=1$ at time $t=0-100$ and $t=0-200$ respectively.}
		\label{fig:wave_phase}
	\end{center}
\end{figure}

We also show the phase space diagram of the largest mode for the two cases 
$\omega=0.5$ and $\omega=1.0$ in Figure~\ref{fig:wave_phase}. 
It is evident from the panels that the wave nature of the trajectory becomes stronger with increasing $\omega $ (Compare Figure~\ref{fig:iso_phase} and~\ref{fig:wave_phase}). One way of quantifying the wave versus turbulent nature of a time series is through analogy to the  signal-to-noise-ratio (SNR) defined as~\cite{Bulsara1996PT,Dmitruk:PoP2009} 
\begin{eqnarray}
SNR = \log_{10}\bigg[ \frac{P(f_0)}{P_0(f_0)} \bigg]\label{snr}.
\end{eqnarray}	
Here $f_0$ is the frequency at the peak (center), corresponding to the applied frequency in the model (Eq.~(\ref{eq:nondim_eq})), and $P_0(f_0)$ is a background value of the power spectrum, if the power law were continued through $f_0$, ignoring the peak at the wave frequency. 
\begin{figure}
	\begin{center}
		\includegraphics[scale=1.0]{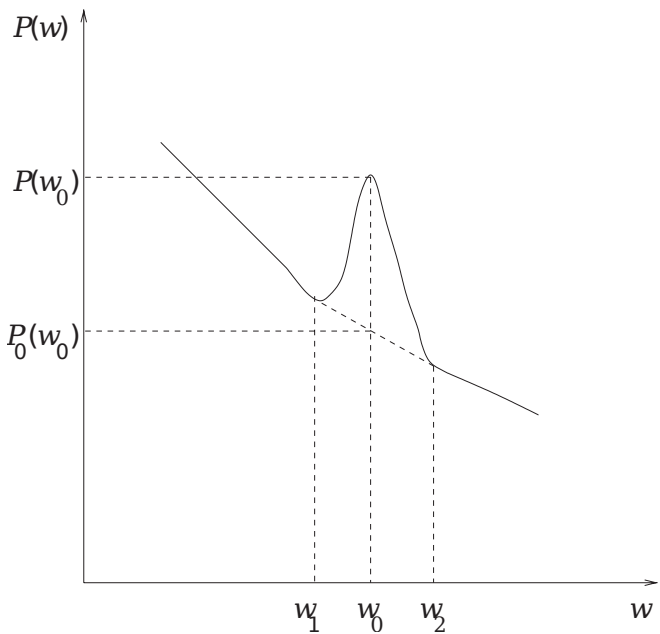}
		\caption{Scheme for illustrating the calculation of signal-to-noise-ratio (SNR). Variable $\mathrm{w}$, used in the figure, and the frequency $f$ are related by $\mathrm{w}=2 \pi f$. (Adapted from~\cite{Dmitruk:PoP2009})}
		\label{fig:snr}
	\end{center}
\end{figure}
The meaning of this parameter is illustrated in Figure~\ref{fig:snr}.
\begin{table}
\caption{SNR for the $k=0.3$ mode for different values of $\omega$.}
	\label{tab:snr}
		\begin{tabularx}{0.5\textwidth}{X X}
			\hline \hline
			$\omega$ 
			& SNR 
			\\
			\hline
			 0.0 & 0.0 \\
			 0.5 & 0.44 \\
			 1.0 & 1.64 \\
			 2.0 & 2.30 \\
			\hline \hline
		\end{tabularx}
\end{table}

The SNR values for the $k=0.3$ mode for different $\omega$, are reported in Table~\ref{tab:snr}. The SNR is $0$ for the $\omega=0$ case and increases monotonically for the increasing values of $\omega$. The fact that the phase space trajectories become more wave-like (See Figure~\ref{fig:iso_phase} \& ~\ref{fig:wave_phase}) with noticable peak at corresponding $\omega$ (See Figure~\ref{fig:1overf_1}~(a)) is reflected through this observation.
This is again consistent with the findings based on 
full MHD simulations \cite{Dmitruk:PRE2007}.
 
\section{\label{sec:discussion}Discussion}
In this paper, we have modified the Langevin equation of Brownian motion and the equation of Kubo-type oscillator to mimic some properties of MHD turbulence. By a systematic conglomeration of von Karman similarity decay hypothesis, Kolmogorov symmetries, and the Langevin process of Brownian motion, the model is shown to produce some results well in agreement with DNS results. A number of similar ``map'' models have been proposed in the past to model hydrodynamic turbulence. Bak \textit{et al.}~\cite{Bak:PRA1988} suggested a discrete ``cellular automata'' may be thought of as a ``toy'' turbulence model. 
Quantitative agreement with experimental results have been shown using Langevin type models (See~\cite{Beck:PhysA1996}). Beck proposed a series of simple cascade models based on an extension of Langevin theory to recover scaling and intermittency 
properties of hydrodynamic turbulence~\cite{Beck:PRE1994, Beck:PhysD1997, Beck:PhysA2001}(Also see~\cite{Hilgers:EPL1999}). Similar Langevin models with complex variables with a linear term have also been used to study simple models for quantum turbulence~\cite{Beck2013PRE,Miah2014EPL}. However, apart from subtle differences, 
our model is different from the earlier models in at least two ways:   
\begin{enumerate}
	\item We do not introduce any cascade in our model. Different length scales evolve independently of each other, while 
 the cascade is modeled by a correlated random forcing.
 \item For the anisotropic case, we introduce a wave term which models the Alfv\'en waves in MHD. This property is absent in incompressible hydrodynamic turbulence.
\end{enumerate}

Due to the structure that we have imposed, the 
present model is able to 
reproduce  a number of important and realistic features of 
MHD turbulence, at extremely low computational cost: 
\begin{itemize}
  \item the model obtains a Kolmogorov inertial range spectrum;
  \item at long wavelengths, modifying 
  a parameter to generate long correlation times, the model obtains 
   a range of modes with $1/f$ frequency spectra;
  \item introduction of wave activity in the model gives rise to expected 
  effects on correlation times, $1/f$ spectra and phase space behavior.
\end{itemize}

While the 
model is not constructed to emulate sweeping time decorrelation, 
it is evident that 
introducing wave activity has the expected effect 
on the time-decorrelations in the computed solutions, in absence of sweeping (See Figure~\ref{fig:aniso_tau-k}) although the energy spectra maintains Kolmogorov scaling in the inertial range even after introducing the wave term (See Figure~\ref{fig:aniso_spectrum}).
  
We remark that this kind of approach is not entirely new in plasma physics and our scheme is similar to that described in TenBarge {\em et al.}~\cite{TenBarge:CPC2014}. Our model validates and generalizes the scheme even further. Plasma systems with  system sizes much larger than kinetic scales show MHD like behavior at large scales but small scales properties like dissipation are described by kinetic physics. Even with state of the art computational resources, PIC kinetic simulations with some of the largest system sizes only begin to probe near the edge of inertial range. The study presented in this paper may be useful to couple kinetic simulations to large scale fluid driving.  
 
The arguments presented here do not capture the sweeping effects properly. The smaller eddies in a turbulent system are swept away by the larger eddies before any significant deformation takes place. As a consequence, a sweeping or advection time scale $(\tau_s \sim k^{-1})$ comes into play apart from the local nonlinear time~\cite{Chen1989PoF} (viscous dissipation time scale becomes important only in dissipation range). This effect induces a profound difference in the Eulerian statistics and the Lagrangian statistics (following fluid elements). Of course the present model, having only a single independent complex degree of freedom at each scale, is too simple to distinguish such Lagrangian effects~\cite{Tennekes1975JFM}. A more complete model would be required to incorporate this level of realism. Whether this can be done with a relatively simple model remains to be established. Further, the arguments presented here do not produce the intermittency effects as observed in MHD. We are in the process of including such effects into a more advanced model like ~\cite{Beck:PRE1994}.

Here we concentrate mostly on turbulent plasma systems. The ideas presented here may be extended to other turbulent systems like convective turbulence, stably stratified flows, rotating turbulence etc. Our approach may be useful to model other chaotic phenomena like stock market exchange, earthquake intensities, prices of commodities etc.
 
\section*{Acknowledgments}
The authors thank Debanjan Sengupta for fruitful discussions during the initial stage of the project, and Aadya Parashar for assistance with the manuscript. The authors are grateful to the two Referees whose inputs helped improve the quality of the manuscript substantially. This research is supported in part by NSF AGS-1063439 and AGS-1156094 (SHINE) and by NASA grants NNX15AB88G (LWS), NNX14AI63G (Heliophysics Grand Challenges), NNX17AB79G (Heliophysics GI Program), and the Solar Probe Plus project under subcontract SUB0000165 from the Princeton University ISOIS project. 


%

\end{document}